\documentclass[a4paper]{jpconf}
\usepackage{graphicx}
\begin{document}
\title{Flat-band ferromagnetism in armchair graphene nanoribbons}

\author{Yen-Chen Lee and Hsiu-Hau Lin}
\address{Department of Physics, National Tsing-Hua University, Hsinchu 30013, Taiwan}
\address{Physics Division, National Center for Theoretical Sciences, Hsinchu 30013, Taiwan}
\ead{hsiuhau@phys.nthu.edu.tw}

\begin{abstract}
We study the electronic correlation effects in armchair graphene
nanoribbons that have been recently proposed to be the building blocks
of spin qubits. The armchair edges give rise to peculiar quantum
interferences and lead to quenched kinetic energy of the itinerant
carriers at appropriate doping level. This is a beautiful
one-dimensional analogy of the Landau-level formation in two
dimensions except the magnetic field is not needed here. Combining the
techniques of effective field theory and variational wave function
approach, we found that the ground state exhibits a new type of
flat-band ferromagnetism that hasn't been found before. At the end, we
address practical issues about realization of this novel magnetic
state in experiments.
\end{abstract}

\section{Introduction}

Graphene\cite{Geim07}, composed carbon atoms arranged in a two-dimensional honeycomb lattice, is the building block for all graphitic materials from the 0D fullerenes to 1D nanotubes and also the common 3D graphite. However, following Peierls and Landau's arguments, the two-dimensional lattice is unstable and cannot exist at any finite temperature. Therefore, graphene is often used as a toy model and viewed as an academic material until its recent discovery in laboratory\cite{Novoselov04}. Since graphene has promising application potentials in nanoscale electronic devices, it is important to study how the band structure changes with respect to the finite transverse width and also the edge topology. In addition, we also expect the electron-electron interaction will play a crucial role for physics properties in low-energy limit.

Therefore, we are motivated to investigate the ground state properties in graphene nanoribbon (GNR) here. With lithography techniques, GNR with width smaller than 20 nm has been fabricated successfully\cite{Han07}. However, the edge roughness seems to post a tough challenge on both theoretical and experimental sides. A breakthrough comes from chemical approach\cite{Li08} recently, that can produce GNR down to 10 nm in the controlled way. Moreover, these GNRs have remarkably smooth edges and make the fast electronics at molecular scale possible\cite{Wang08}. In this work, we concentrate on the GNR with armchair edges since the open boundaries give rise to peculiar flat bands. Note that the edges are hydrogenated so that the dangling $\sigma$ bonds are saturated and only the $\pi$ bands remain active in low energy\cite{Son06}.

By combining analytic weak-coupling analysis, numerical density matrix renormalization-group (DMRG) method, and the first-principles calculations, it was shown\cite{Lin08} that armchair GNR exhibits novel carrier-mediated ferromagnetism upon appropriate doping. Even though only $\pi$ bands are active in low-energy, in appropriate doping regimes, the armchair edges give rise to both itinerant Bloch and localized Wannier orbitals. These localized orbitals are direct consequences of quantum interferences in armchair GNR and form flat bands with zero velocity. The carrier-mediated ferromagnetism can thus be understood in two steps: Electronic correlations in the flat band generate intrinsic magnetic moments first, then the itinerant Bloch electrons mediate ferromagnetic exchange coupling among them. Here we use a variational wave function approach and try to understand how the flat-band ferromagnetism evolves with the interaction strength.

\section{Hubbard Model for Armchair Graphene Nanoribbon}

We start with the Hubbard model to describe the armchair GNR,
\begin{eqnarray}
H=H_{t}+H_{U} 
=-t\sum_{\langle{\bf r},{\bf r'}\rangle,\alpha}[c_{\alpha}^{\dag}({\bf r})c_{\alpha}({\bf r'})+h.c.]
+U\sum_{{\bf r}} n_{\uparrow}({\bf r})n_{\downarrow}({\bf r}).
\end{eqnarray}
where $t$ is the hopping amplitude, $U>0$ is the repulsive on-site interaction, $\alpha=\uparrow,\downarrow$ is the spin index, ${\bf r=(x,y)}$, and $\sum_{\langle{\bf r},{\bf r'}\rangle}$ is taken only for nearest neighbor bonds. Rough estimates from experiments give $t\simeq3$ eV and $U\simeq6$-$10$ eV, putting the ratio $U/t$ to be of order one.

Before diving into the numerical details, it is insightful to highlight the peculiar flat-band orbitals first. For the armchair GNR, its transverse width can be labeled by an integer $L_y$, i.e. the size of the unit cell for each sublattice. Writing down the wave function on different sublattices as the two-componenet spinor, it is easy to check that, for symmetric armchair GNR (odd $L_y$), one can construct a localized wave function at $x=x_0$,
\begin{eqnarray}
\Psi_{F\pm}({\bf r}) = \delta_{x,x_0} \frac{1}{\sqrt{L_y+1}} \left[
\begin{array}{c}
\sin (\pi y/2)\\
\mp \sin (\pi y/2)
\end{array}\right],
\end{eqnarray}
and show it is indeed an eigenstate of the hopping Hamiltonian. Applying translational invariance to shift the Wannier orbital, the flat band emerges at the end. This is the one-dimensional analog of the Landau level degeneracy for two-dimensional electrons in magnetic field. It is interesting that the edge topology in 1D quenches the kinetic energy without the necessity to couple to external magnetic field. Furthermore, we would like to point out that the flat-band ferromagnetism in armchair GNR is not the same by Mielke-Tasaki\cite{Mielke93,Tasaki98} mechanism since there is no direct overlap between these flat-band orbitals. The ferromagnetism is actually mediated by other dispersive bands that couple the magnetic moments in the flat band.

\section{Variational Energy}

On the purpose of seeing the polarization for the ground state, we introduce the following trial wave function,
\begin{eqnarray}\label{try}
|\Psi_0\rangle = \prod_{s,m}
\left[\prod_{k_{s m \uparrow}} a_{s m \uparrow}^{\dag}(k_{s m \uparrow}) \right]
\left[\prod_{k_{s m \downarrow}} a_{s m \downarrow}^{\dag}(k_{s m \downarrow}) \right]
|0\rangle,
\end{eqnarray}
where $s=\pm1$ denotes the valence bands ($V$) and the conducting bands ($C$) separately and $m=1,2,..L_{y}$ is the band index. Thus, there are $2L_{y}$ number of bands totally. The lower bounds for $k_{s m\alpha}$ are decided by their band structures (which are either $0$ or $\pm\pi$), while the upper bounds are the fitting parameters to minimize the variational energy, reducing the problem to look for minimum in the $4L_{y}-1$-dimensional (one constraint from the total particle conservation) parameter space. For convenience, we introduce the filling factor, $0< \nu_{sm\alpha} <1$ for each band as the variational parameters. Since the filling factor denotes the fraction of the filled states in the band, summing $\nu_{sm\alpha}$ over all bands gives the number of particles in one unit cell, i.e. $2L_{y} \langle n \rangle$, where $0< \langle n \rangle < 2$ is the average particle number per site. Meanwhile, the total spin of the ground state, $S=\frac{1}{2}\sum_{s,m}(\nu_{s m\uparrow}-\nu_{s m\downarrow})$, can also be expressed in terms of the filling factors easily.

The kinetic energy per unit cell is obtained by integrating the energy spectra to the desired fillings. The integration is fundamental and can be expressed in terms of the elliptic integrals,
\begin{eqnarray}
E_{t} =
\sum_{sm\alpha}st \left(\frac{4\gamma_{m}}{\pi}\right)
\left[E\left(\frac{k^f_{sm\alpha}}{4},\frac{8\cos q_{m}}{\gamma_{m}^{2}}\right)-
E\left(\frac{k^i_{sm\alpha}}{4},\frac{8\cos q_{m}}{\gamma_{m}^{2}}\right)
\right]  
+\sum_{s\alpha}st\nu_{s_{F\alpha}},
\end{eqnarray}
where $E(\phi,\eta)=\int^{\phi}_{0}(1-\eta\sin^{2}\theta)^{1/2}d\theta$ for $-\pi/2<\phi<\pi/2$ is the elliptic integral of the second kind. The magnitude of the transverse momentum is $q_m = m\pi/(L_y+1)$ and $\gamma_{m}=1+2\cos q_{m}$. For clarity, we separate the kinetic energy from the dispersive bands and the flat bands in above. Thus, the summation of the band index $m$ does not include the flat bands. It shall be clear that the kinetic energy of the flat band (denoted by $F$) is rather trivial since it is directly proportional to filling factor multiplied by the energy $\pm t$. For dispersive bands, the initial momentum $k^i_{sm\alpha}$ starts from either $0$ or $\pi$, depending where the spectrum minimum sits in the 1D Brillouin zone. The corresponding filled momentum $k^{f}_{sm\alpha}$ is either $\pi \nu_{sm\alpha}$ or $\pi (1-\nu_{sm\alpha})$.

Similarly, we can also compute the interaction energy per unit cell,
\begin{eqnarray}
E_{U} = \sum_{sm} \sum_{s'm'}
\left[\sum_{y} \frac{U}{(L_{y}+1)^{2}} \sin^2(q_m y) \sin^2(q_{m'}y) \right]
\nu_{sm\uparrow} \nu_{s'm'\downarrow},
\end{eqnarray}
where the effective repulsive interaction $U(sm,s'm')$ between the bands $sm$ and $s'm'$ is the summation inside the bracket in above. The reason why the effective repulsion does not carry the explicit dependence on $s, s' = \pm 1$ is due to the particle-hole symmetry in the armchair GNR. The analytic expressions for the kinetic and interaction energies are nice but the search for minima in the global parameter space can only be done by numerical approach due to the number of all filling factors are large.

\section{Numerical Results}\label{sec}

\begin{figure}
\centering
\includegraphics[width=14pc]{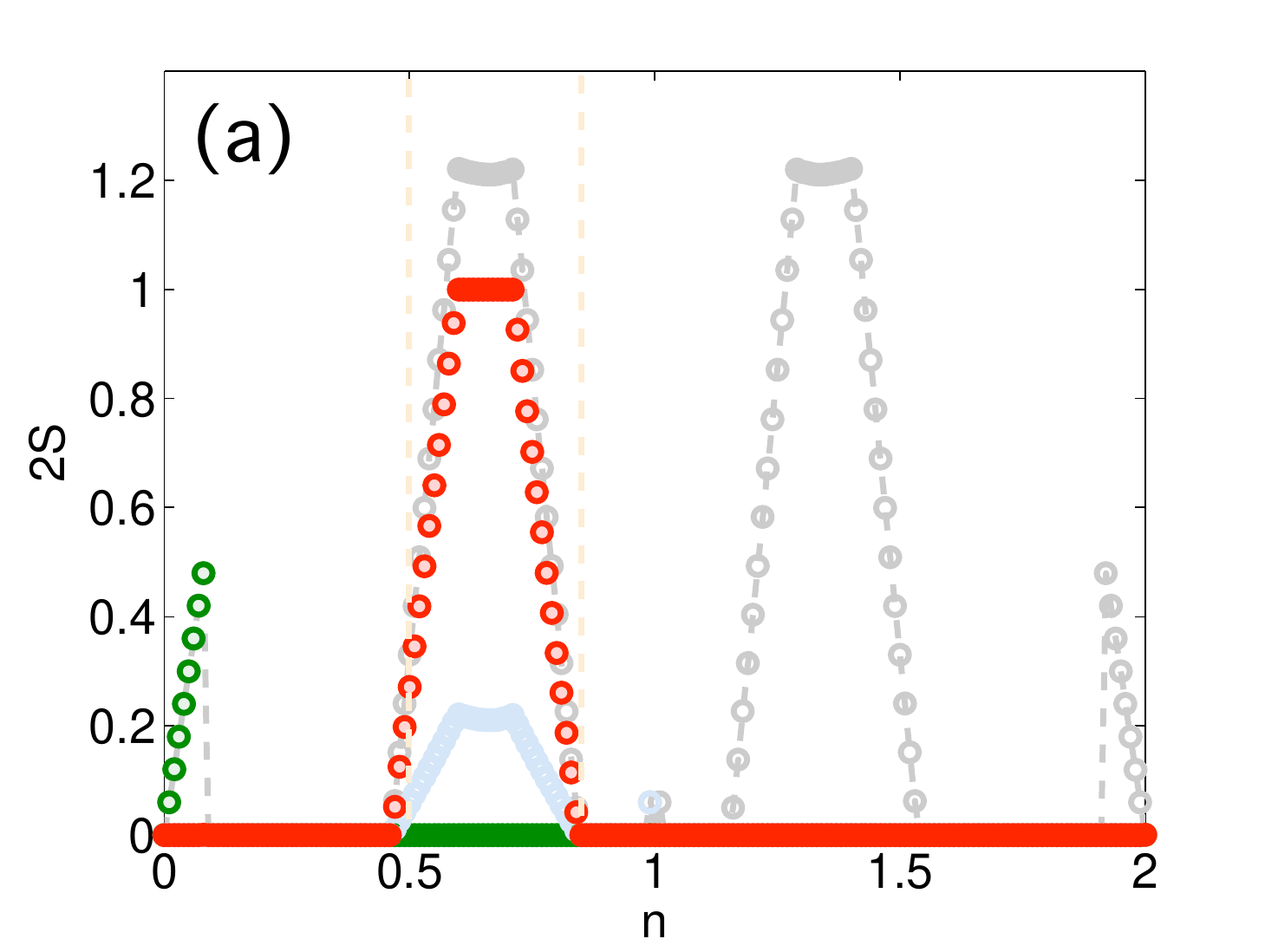}
\hspace{2pc}
\includegraphics[width=14pc]{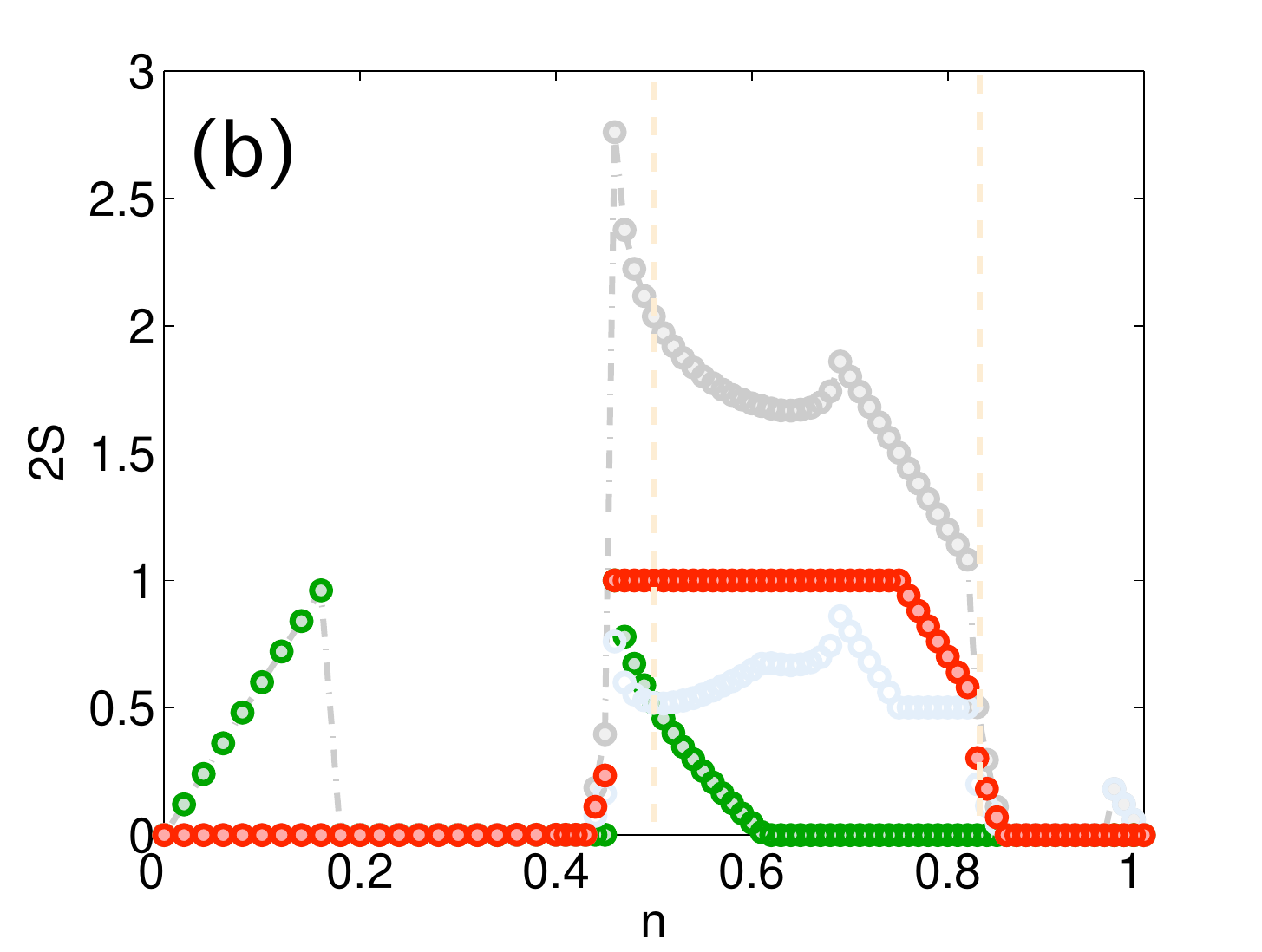}
\caption{\label{fig:S} Spin polarization in each bands at different electron number per site $n$ for the (a) $u=2$ and (b) $u=4$ cases. The flat-band regime ($1/2 < n < 5/6$) is between the yellow dashed lines and the gray open circle denotes twice of the total spin per unit cell $2S$. Spin polarization from the valence bands are shown in red, green, blue open circles, corresponding to the flat band, the lower ($m=1$) and the upper ($m=3$) dispersive bands. The spin polarization for the conduction bands are mirror images of these around the symmetric axis $n=1$. For clarity of the figure, they are not shown here.}
\end{figure}

The numerical search is carried out extensively at interaction strength from $u \equiv U/t =1$ to $u=7$. As an illustrating example, we present our numerical results at $u=2$ and $u=4$ in Fig.~\ref{fig:S} for $L_y=3$ armchair GNR. For wider GNR, the numerical results are qualitatively the same. Meanwhile, to avoid being trapped in local minima, we randomly choose the initial searching points and select the final configurations with lowest energy. 

For the relatively weak interaction strength $u=2$, the influence of the band structure is still manifest. First of all, the particle-hole symmetry is manifest in the total spin $S$. NOte that the finite spin polarization near $n=0$ and $n=2$ is due to the Van Hove singularity and is not our main focus here. It is rather interesting that the ferromagnetic phase coincide with the flat-band regime rather nicely. In fact, the total spin per unit cell $S$ mainly comes from the flat-band orbitals. Furthermore, since the upper dispersive band ($m=3$ in light blue color) intersects with the flat band, it is also polarized in the flat-band regime. On the other hand, the lower dispersive band ($m=1$ in green color) is separated by a finite gap and does not participate in the ferromagnetism. These numerical findings agree with field-theory approach, which shows that the exchange coupling between the flat-band moments are mediated by the itinerant carriers in the intersecting dispersive bands.

As the interaction strength increases to $u=4$, the competition between the kinetic and the interaction energies become highly non-trivial even at the mean-field level. The ferromagnetic regime due to Van Hove singularity expands a little bit as expected from Stoner criterion. It is rather remarkable that the ferromagnetic phase is still confined to the flat-band regime rather well. But, since the interaction strength is large enough, both dispersive bands contribute despite the minor gap in one of them. The intersecting band greatly enhances the ferromagnetic correlations and accounts for roughly one-third of the total spin density. This implies that the itinerant carriers no longer play the secondary role as messengers to line up the localized moments in the flat band. Instead, they also develop ferromagnetic correlations and contribute to the total spin density. The role of the dispersive band with a minor gap is peculiar. It participates the ferromagnetic phase in the first-half of the flat-band regime and drops out in the second-half. It is not yet clear whether this asymmetrical behavior is an artifact of the simple wave function we chose, or it hints for something real in realistic armchair GNR.

\section{Summery}

We show that the armchair edges of GNR quenches the kinetic energy of the itinerant
carriers and leads to the flat-band orbitals at finite doping. This is a beautiful one-dimensional analogy of the 2D Landau levels except the magnetic field is not needed here. Following the hints from the field-theory analysis, we study magnetic properties of the ground state via variational wave function approach and found the flat-band ferromagnetism in armchair GNR. We acknowledge support from the National Science Council of Taiwan through grants NSC-95-2112-M-007-009 and NSC-96-2112-M-007-004 and also from the National Center for Theoretical Sciences in Taiwan.

\end{document}